\documentclass[aps,prl,twocolumn]{revtex4-1}

\usepackage{latexsym}
\usepackage{slashed}
\usepackage{dcolumn}   
\usepackage{bm}        
\usepackage{amsmath}
\usepackage{amssymb}   
\usepackage{longtable}
\usepackage{float}
\newcommand{\be}{\begin{equation}}
\newcommand{\ee}{\end{equation}}
\newcommand{\bea}{\begin{eqnarray}}
\newcommand{\eea}{\end{eqnarray}}
\newcommand{\bma}{\begin{matrix}}
\newcommand{\ema}{\end{matrix}}

\newcommand{\bml}{\begin{mathletters}}
\newcommand{\eml}{\end{mathletters}}
\newcommand{\bes}{\begin{subequations}}
\newcommand{\ees}{\end{subequations}}

\newcommand{\bi}{\begin{itemize}}
\newcommand{\ei}{\end{itemize}}
\newcommand{\gev}{~{\rm GeV}}
\newcommand{\tev}{~{\rm TeV}}

\begin{document}
\title{Topologically stable, finite-energy electroweak-scale monopoles}
\author{P. Q. Hung}
\email{pqh@virginia.edu}
\affiliation{Department of Physics, University of Virginia,
Charlottesville, VA 22904-4714, USA}

\date{\today}

\begin{abstract}
The existence of a magnetic monopole, if it exists, remains elusive. Experimental searches have been carried out and are continuing in this quest. Of great uncertainty is the mass of the monopole which is model-dependent and ranging from some Grand Unified scale to the electroweak scale. In this paper, we propose a model where topologically stable, finite-energy monopoles {\em \`{a} la} 't Hooft-Polyakov could exist with a mass proportional to the electroweak scale. This comes about in a model of neutrino masses where right-handed neutrinos are {\em non-sterile} whose electroweak-scale Majorana masses are obtained by the coupling to a complex {\em triplet} Higgs field. Custodial symmetry which insures $M_W=M_Z \cos \theta_W$  requires the introduction of another triplet Higgs field but {\em real} this time. It is this {\em real Higgs triplet} that is at the core of our proposal.

\end{abstract}

\maketitle

\section{Introduction}
The fascinating idea of a magnetic monopole has been around since the time when Dirac was intrigued by why electric charges are quantized. Dirac's postulate of a {\em point-like} magnetic monopole with a semi-infinite "singular string" attached to it. The wave function for an electron going along a closed loop encircling the singular string is single valued if the Dirac quantization condition $eg_{D} =n/2$ is satisfied with $g_D$ being the magnetic charge. Another way to say this is that, in order for the string to be undetected, the quantization condition has to be satisfied. Electric charges are quantized in units of $1/2g_D$ in the presence of a Dirac point-like monopole.

't Hooft and Polyakov \cite{thooft,polyakov} discovered that, by embedding the "electromagnetic" $U(1)_{em}$ into the Georgi-Glashow(G-G) \cite{GG} gauge group $SO(3)$, there exists a stable, finite energy monopole-like solution where the singular string can be gauge-rotated away. The 't Hooft-Polyakov monopole necessitates the existence of a {\em real} Higgs triplet of $SU(2)$. However, the G-G model as it stands is not the correct electroweak model as we know it. It was subsequently realized that, by embedding the SM into a Grand Unified (GUT) gauge group such as $SU(5)$, one could construct topologically stable, finite-energy (TSFE) magnetic monopoles which are extremely heavy with a mass proportional to the GUT scale over the GUT fine structure constant \cite{preskill}. 

One could also ask the question of whether or not the Standard Model (SM) could contain such a monopole with a mass which is now proportional to the electroweak scale and which could, in principle, be produced and detected. Unfortunately, it is well-known that the SM which contains only Higgs doublets does not have TSFE monopoles by topological arguments for the $SU(2)$ part, as we shall briefly review below. An alternative way out of this conundrum is a proposal of Cho and Maison \cite{cho-maison}which asserted that the SM monopole can exist by looking considering the SM $SU(2) \times U(1)$ as a gauged $CP^1$ model with the Higgs doublet as a $CP^1$ field and it is this Higgs field that admits a topologically stable monopole configuration. The Cho-Maison electroweak monopole however suffers from a divergence when one tries to compute its mass classically. Remedies to that problem, essentially by modifying the kinetic energy term of the $U(1)_Y$ gauge field, have been proposed \cite{cho-maison2} at the price of a non-negligible uncertainty in the monopole mass and the introduction of some unknown physics at energies above the electroweak scale. Despite that uncertainty, it is a very plausible way to generate electroweak monopoles (and dyons) in the SM, a possibility with great experimental implications. 

However, one could also ask whether or not it is possible to have a monopole solution in a generalized SM with an extended Higgs sector (and extended fermion sector as in the model of \cite{pqnur}) where a real Higgs triplet exists. The question is as follows: under what condition(s) would such a monopole exist when both Higgs doublet and triplet breaks the electroweak symmetry? In the next section, we wish to show that electroweak monopoles coming from the real triplet can exist even in the presence of Higgs doublets. Furthermore, we wish to present two possible scenarios but will concentrate on one of them in this paper: 1) Assuming the existence of some unknown physics beyond the electroweak scale, one can have monopole solutions coming from both the real triplet and the doublet; 2) Assuming no new physics beyond the electroweak scale (at least of the type that can modify the $U(1)_Y$ gauge kinetic term), the only topologically stable, finite energy monopole will be one that comes from the real triplet. It is the latter possibility that we will concentrate on in this paper. The first possibility is also very interesting and will be explored in a separate paper.

In what follows, we will restrict ourself to the SM gauge group $SU(3) \times SU(2) \times U(1)$ with no additional gauge interactions. We will address the physical motivation for having a real Higgs triplet, in addition to the usual doublet in the SM. Below is a brief description of a physically-motivated model where the introduction of a real Higgs triplet was needed to preserve the so-called Custodial Symmetry.

In this manuscript, we propose a model which contains topologically stable, {\em finite-energy} electroweak-scale monopoles {\em \`{a} la} 't Hooft-Polyakov. More precisely speaking, this is not a model proposed for the electroweak monopole but rather this solution is a consequence of a model proposed for something else: the possible existence of {\em non-sterile} right-handed neutrinos with electroweak-scale masses in a seesaw mechanism for light neutrinos \cite{pqnur}. The seesaw mechanism within the framework of this model can be tested directly at colliders by searching for like-sign dileptons at displaced vertices, among many other collider signals. The main reason this model (EW-$\nu_R$ model) can give rise to electroweak monopoles is as follows. As \cite{pqnur} has described, right-handed neutrinos acquire electroweak-scale Majorana masses $M_R \propto \Lambda_{EW} \sim 246 \gev$ by coupling to a {\em complex} Higgs triplet $\tilde{\chi}$. The Z-width requires that $M_R \geq 46 \gev$ since $\nu_R$'s are non-sterile, which translate into $\langle \tilde{\chi} \rangle=v_M \propto \Lambda_{EW}$. This however will {\em badly} destroy the well-known and experimentally successful relationship $M_W=M_Z \cos \theta_W$, a consequence of the so-called Custodial Symmetry, {\em unless} another Higgs {\em real} triplet $\xi$ exists with $\langle \xi \rangle = \langle \tilde{\chi} \rangle=v_M$ \cite{triplet}. As we shall see below, it is this real triplet $\xi$ which gives rise to the electroweak monopole solution. (For more details on the phenomenology of the EW-$\nu_R$ model, please consult \cite{pqnur,pqnur2,pqnur3}.) The logical chain of our arguments is as follows: $M_R \propto \Lambda_{EW} \rightarrow$ Complex triplet $\tilde{\chi}$, Custodial symmetry $\rightarrow$ Real triplet $\xi \rightarrow$ topologically stable, {\em finite-energy} electroweak-scale monopoles. This chain shows the deep connection between neutrino masses and the possibility of electroweak monopoles.  

Since the core of the arguments given below involves two Higgs triplets (one complex, $\tilde{\chi}$, and one real, $\xi$), it is legitimate to ask whether or not there exists Nambu-Goldstone (NG) bosons. The answer is {\em no} as explained in \cite{pqnur,pqnur2}. In a nutshell, it has to do with the fact that the model contains a global symmetry which would have given NG bosons when it is spontaneously broken but this symmetry is explicitly broken by terms in the potential which are required for a proper vacuum alignment. 

For reasons to be given below, we shall call the topologically stable, {\em finite-energy} electroweak-scale monopoles by the name: {\bf $\gamma$-Z magnetic monopole}. 

We will show that the EW-$\nu_R$ model contains very interesting non-perturbative solutions in the form of $\gamma$-Z magnetic monopoles upon a close examination of the global structure of the model. This provides another (non-perturbative) characteristic signal to be searched for at dedicated experiments such as MoEDAL and CMS,ATLAS, LHCb among others. As we will see below, the $\gamma$-Z magnetic monopole is a finite-energy soliton whose mass is concentrated in a core of radius $\sim 1/M_Z$. This mass is expected to be $\sim (4\pi v_M/g \times O(1)$, where $g$ (instead of $e$) is the $SU(2)$ coupling for reasons to be explained below. With $(\sum_{i=1,2} v_i^2 + v^{M,2}_{i}) + 8 v_M^2 = (246 \gev)^2$, one obtains $46 \gev < v_M< 87 \gev$, the $\gamma$-Z magnetic monopole mass is roughly between 900 GeV and 3 TeV. The long-range magnetic field appearing outside the core behaves like $(1/g)(1/r^2)=(1/e)(\sin \theta_W)(1/r^2)$. 

The plan of the paper is as follows. We first summarize the Cho-Maison construction of the monopole in the SM with a Higgs doublet. We then discuss the inclusion of an additional real Higgs triplet and the condition under which a 't Hooft-Polyakov monopole solution can be realized. This "toy model" is used strictly to discuss the aforementioned monopole issue. A physically-motivated model where this real triplet is present will be discussed following that prelude. Under the requirement of finite energy for the monopole solution with no new physics which could modify the gauge kinetic term, we discuss properties of this monopole within the framework of the EW-$\nu_R$ model \cite{pqnur}.

\section{A brief review of monopoles in the SM with a Higgs doublet and its extension to a real Higgs triplet}

To set the tone for what follows in the next few sections, we briefly review the construction of the Cho-Maison electroweak monopole in the SM. Since we will review some results of homotopy theory below, we will simply quote the essential one in this section: the non-triviality of the second homotopy group i.e. $\pi_2 \neq 0$ is a requirement to have a topologically stable monopole. If one only considers $SU(2)$, the vacuum manifold of the complex Higgs doublet is represented by a 3-sphere $S^3$ and homotopy tells us that $\pi_2(S^3)=0$. There is {\em no} topologically stable monopole in the SM with only Higgs doublets. There is however a twist to this argument as shown by \cite{cho-maison}: the Higgs doublet field can be viewed as a $CP^1$ field of a gauged $CP^1$ model when $U(1)$ is taken into account and now $\pi_{2}(CP^1 \cong S^2)=Z$. Hence, there is a topologically stable monopole (and dyon). However, the computation of the energy of the monopole in order to estimate its mass yields an infinite quantity which necessitates an ad-hoc modification of the SM. In what follows we will summarize some of the essential points of the Cho-Maison construction, leaving out details which could be obtained from \cite{cho-maison}. 

The Cho-Maison spherically-symmetric ans{\"a}tz for the Higgs doublet, $\phi = \frac{1}{\sqrt{2}}\rho \xi$ ($\xi^{\dagger} \xi=1$),  is written as 
\be
\label{CM}
\rho=\rho(r) ; \xi=\imath \left(\begin{array}{c}
	   \sin(\theta/2) e^{-\imath \varphi} \\
	   -\cos(\theta/2) \\
	  \end{array}\right) \,.
\ee
From the above ans{\"a}tz, one can write down the equation of motion and solve it numerically \cite{cho-maison}. Of importance are the boundary conditions: 1) $\rho \rightarrow c \, r^{\delta}$ as $r \rightarrow 0$ ($\delta=(\sqrt{3}-1)/2$); 2) $\rho \rightarrow \rho_0 + \rho_{1} \frac{\exp (-\sqrt{2} \mu r)}{r}$ as $r \rightarrow \infty$ with $\rho_0 = v$. The computation of the monopole energy which would determine its mass yields two terms, one of which is {\em infinite}, namely $E_{0} =\frac{1}{4 \pi} \int^{\infty}_{0} \frac{dr}{2 r^2}\{\frac{1}{g^{'2}} + \frac{1}{g^2}(f^2-1)^2 \}$. Since the existence of this topologically stable monopole, albeit one with an infinite energy, in the SM with a Higgs doublet comes about due to the existence of the $U(1)$ hypercharge gauge group, it was realized \cite{cho-maison2} that a {\em modification} of the SM was needed, in the form of a permittivity factor in front of the $U(1)$ gauge kinetic term $\frac{\epsilon (|\phi|/v)}{4}B_{\mu \nu} B^{\mu \nu}$ with $\epsilon (|\phi|/v)$ going to zero as $r \rightarrow 0$ and to 1 as $r \rightarrow \infty$. Basically, the $U(1)$ gauge coupling $g^\prime$ is rescaled to $\bar{g}^\prime = g^\prime/\sqrt{\epsilon}$ which goes to infinity as $\epsilon \rightarrow 0$ when $|\phi|$ vanishes at the origin. This modification of the SM in order to have a finite-energy solution would obviously come from some unknown dynamics beyond the electroweak scale. 

It is worth mentioning that the magnetic charge of the Cho-Maison monopole is given by $q_m = 4 \pi/e$, twice the value of that of a Dirac monopole.

At this point, one might be tempted to ask whether or not a 't Hooft-Polyakov monopole could exist if one introduces a {\em real} Higgs triplet, in addition to the doublet. The physics motivation and the constraint on such a triplet will be discussed in the next section. To be consistent with the notation used in \cite{pqnur} and below, we will denote such real triplet by $\xi$, not to be confused with $\xi$ used in Eq.~(\ref{CM}). As we shall see in the next section, the 't Hooft-Polyakov ans$\ddot{a}$tz for $\xi$ is given by $\xi^a = \frac{r^a}{g r^2} H(v_M g r)$. Asymptotically, $\rho \rightarrow \rho_0=v_2$ and $\xi^a \rightarrow v_M r^a/r$ as $r \rightarrow \infty$. 
Numerically, for a certain range of parameters, both $\rho(r)$ and $H(r)$ approach their asymptotic values fairly rapidly. Specifically, as shown in \cite{cho-maison}, for the Higgs doublet self-coupling $\lambda_{2}/g^2=1/4$, $\rho / \rho_0 \rightarrow 1$ when $r \sim 3/(gv_2/2)=6/(gv_2)$. (If there were only a single doublet, $gv_2/2$ would simply be $M_W$.) From \cite{shnir}, one can take as an example the triplet self-coupling $\lambda_3=1$ and one can see that $H(v_M g r) \rightarrow 1$ for $r \sim 2/(gv_M)$.
Both types of monopoles could exist if $v_M \sim O(v_2)$ which, in fact, is the case for the EW-$\nu_R$ model \cite{pqnur}. This condition should apply even to the case where there is no dynamics that can regularize the energy of the Cho-Maison monopole.

It is possible that both types of monopoles could exist if there is some dynamics beyond the electroweak scale which could modify the SM with the type described above. It would be interesting to see if such a possibility could exist.

In what follows, we will require that the monopole solution yields  a topologically stable, finite-energy electroweak monopole without the help of unknown dynamics beyond the electroweak scale. We will therefore concentrate on the 't Hooft-Polyakov-type monopole with a real Higgs triplet.

\section{Global structure of the EW-$\nu_R$ model and the $\gamma$-Z magnetic monopole }


Since the 't Hooft-Polyakov monopole is crucial in our subsequent discussion, we will come back to it after a brief excursion into how topologically stable monopoles arise from consideration of results of homotopy groups of spheres.

To find a finite-energy field configuration at spatial infinity which corresponds to a monopole, one requires that the Higgs field approaches its minima, the so-called vacuum manifold, which forms a sphere in 3-dimensional internal space denoted by $S^2$. That is one maps a 3-dimensional spatial sphere to the sphere of vacuum manifold $S^2$. In homotopy theory, this amounts to the second homotopy group $\pi_2$ (for 3-dimensional space) and $\pi_{2}(S^2)=Z$, where $Z=0,1,2,..$. First, $Z$ or simply $n=0$, the {\em winding number}, corresponds to the trivial vacuum with no monopole while $n=1$ corresponds to the first non-trivial solution and so on. The monopole solution is {\em topologically stable} (i.e. the Higgs vacuum manifold forms a 2-sphere $S^2$) because of the fact that it takes an infinite amount of energy to go from the configuration $n=1$ to $n=0$ for example. An explicit example is the Georgi-Glashow model $SO(3) \sim SU(2)$ with a {\em real} Higgs triplet $\xi=(\xi_0, \xi_1, \xi_2)$. In this model, the vacuum manifold is $\xi_0^2+\xi_1^2+\xi_2^2=v_M^2$ corresponding to $S^2$ and the model can accommodate a topologically-stable monopole. 

Can $SU(2)$ accommodate topologically-stable monopoles with different Higgs representations? We are particularly interested in a situation in which $SU(2)$ contains, beside the real triplet $\xi$, a complex triplet $\chi$ and complex doublets $\phi_i$ such as the case with the EW-$\nu_R$ model \cite{pqnur}.
A summary of some of the homotopy theory results is in order here.
\be
\label{pi1}
\pi_n(S^n)=Z; 
\ee
\be
\label{pi2}
\pi_i(S^n)=0 , i<n;  
\ee
\be
\label{pi3}
\pi_n(S_1 \times S_2 \times ...\times S_k)= \pi_n(S_1) \oplus \pi_n(S_2) \oplus ... \oplus \pi_n(S_k),
\ee
where $S_1,..,S_k$ denote spaces which, in our cases, represent different vacuum manifolds. (For pedagogical purposes, what is usually meant by a n-sphere $S^n$ is simply $x_1^2 +x_2^2 +..+x_{n+1}^2 = constant$ so a 2-sphere is $x_1^2+x_2^2+x_3^2= constant$.)

The vacuum manifold of the SM with only a complex Higgs doublet (four independent degrees of freedom) is represented by $ \phi_1^2 + \phi_2^2 + \phi_3^2 + \phi_4^2 = v^2$. This is a 3-sphere $S^3$. From the above results, one has $\pi_2(S^3)=0$ and the SM has {\em no topologically stable monopoles}, a well-known result. This is true for {\em any} number of complex Higgs doublets. A complex Higgs triplet $\tilde{\chi}=(\chi^0, \chi^+, \chi^{++})$ has {\em six} real components and the vacuum manifold $\sum_{i=1}^{I=3} (Re \chi^2_i + Im \chi^2_i)=v_{M}^2$ is represented by a 5-sphere $S^5$. One has $\pi_2(S^5)=0$. As we have stated above, a real Higgs triplet ($\xi$) vacuum manifold is represented by  2-sphere $S^2$ and $\pi_2(S^2)=Z$.

The EW-$\nu_R$ model has the following Higgs content: 1) One real triplet $\xi$; 2) One complex triplet $\tilde{\chi}$; 3) Four complex Higgs doublets, $\phi^{SM}_{i}$ and $\phi^{M}_{i}$ with $i=1,2$, which couple to the SM and mirror fermions respectively. (It also contains Higgs singlets $\phi_S$ which are important for different reasons but are not relevant here.) (Notice that the proper vacuum alignment which guarantees the so-called custodial symmetry gives $\langle \tilde{\chi} \rangle= \langle \xi  \rangle= v_M$ \cite{pqnur}.) The vacuum manifolds of that Higgs sector is 
\be
\label{vacuum}
S_{vac} = S^{2} \times S^{5} \times \sum_{i=1,2} S^{3}_{SMi} \times \sum_{i=1,2} S^{3}_{Mi}  \,.  
\ee  
From (\ref{pi3}), one obtains the second homotopy group of the vacuum manifold of the EW-$\nu_R$ model (\ref{vacuum}) as
\bea
\label{result}
\pi_2(S_{vac})&=&\pi_2(S^2)  \oplus \pi_2(S^5)  \oplus_{i=1,2} \pi_2(S^{3}_{SMi,Mi}) \\ \nonumber
                       &=&  \pi_2(S^2) = Z \,.
\eea
The EW-$\nu_R$ model {\em can accommodate a topologically stable monopole} because of the existence of the real $SU(2)$ triplet $\xi$! (It is amusing to note that custodial symmetry requires $S^2$ (related to $\xi$) has the same "radius" $v_M$ as that $S^5$ (related to $\tilde{\chi}$).)

In what follows, the treatment of the monopole in the EW-$\nu_R$ model follows that of the 't Hooft-Polyakov monopole, except for the final expression of the magnetic field as we shall see below. Last but not least, let us again recall that the SM with {\em only Higgs doublets} has {\em no finite-energy monopoles} for two reasons: 1) $\pi_2(S^3)=0$; 2) The unbroken subgroup of $SU(2) \times U(1)_Y$ being $U(1)_{em}$ can only accommodate a Dirac monopole which is singular at the origin. 
 
The next two steps that we would like to make is to first write down the 't Hooft-Polyakov ans$\ddot{a}$tz and estimate the mass and size of the monopole. Next, we look at the perturbative spectrum by performing {\em small fluctuations} around the background of the 't Hooft-Polyakov solution. In particular, we would like to see how this perturbative spectrum interacts with the monopole. We first start with the case with only the real triplet $\xi$ and include the other scalars (complex triplet $\chi$, doublets $\phi_i$) as parts of the perturbative spectrum. In this first step, $\langle \xi \rangle$ induces $SU(2)_W \times U(1)_Y \rightarrow U(1)_W \times U(1)_Y$. This is where the t Hooft-Polyakov monopole enters our model. As we have argued above, this is also where the only non-perturbative solutions exist since other Higgs representations present in the model (a complex triplet and Higgs doublets) have no topologically stable monopole solutions. 



1) The 't Hooft-Polyakov ans$\ddot{a}$tz is given by
\be
\label{ansatz}
\Xi^a =\frac{r^a}{g r^2} H(v_M g r) ; W^{a}_{n} = \epsilon_{aji}\frac{r^j}{g r^2}[1-K(v_M g r)] ; W^{a}_{0}=0 \,,
\ee
where $a=1,2,3$ and $n=1,2,3$ are the group and space indices respectively. In (\ref{ansatz}), we use $\Xi$ and $W$ to denote the non-perturbative solutions and we shall use the lower cases to denote the perturbative spectrum: $\xi$ and $w$. The boundary conditions at infinity for $H$ and $K$ are $H \rightarrow v_{M } \,g \, r$ and $K \rightarrow 0$ as $r \rightarrow \infty$. At $r=0$, $K-1 \rightarrow 0$ and $H \rightarrow 0$. The differential equations for $H$ and $K$ are well-known in the literature and will not be repeated here. It suffices to state that only numerical solutions are known.They depend on the self-coupling of the scalars and were found to rapidly approach their asymptotic values for sufficiently large values of that coupling.

2) The next question concerns charge quantization far away from the core of the $\gamma$-Z monopole.  As we have mentioned above, $\langle \xi \rangle$ induces $SU(2)_W \times U(1)_Y \rightarrow U(1)_W \times U(1)_Y$. The quantization condition and the associate monopole come from $SU(2)_W  \rightarrow U(1)_W$ and necessarily involves the weak isospin coupling $g$. 

There are two ways to look at charge quantization in our model. The first way is a topological argument and the second way is the requirement that the wave function of a charged particle such as the electron is single-valued as it goes around the monopole.

The topological argument is similar to the 't Hooft-Polyakov and has been discussed at length in the literature. Here, we just summarize the key points. The field strength corresponding to $U(1)_W$ (in $SU(2)_W \times U(1)_Y \rightarrow U(1)_W \times U(1)_Y$) is $W_{3}^{\mu \nu}=\partial^{\mu} W_{3}^{\nu}-\partial^{\nu} W_{3}^{\mu} + \frac{1}{v_{M}^{3} g} \varepsilon_{abc} \xi^{a} \partial^{\mu} \xi^{b} \partial^{\nu} \xi^{c} $. One can then construct a {\em topological current} $k_\mu = \frac{1}{2} \epsilon_{\mu \nu \sigma \rho} \partial^{\nu} W_{3}^{\sigma \rho}$ which is automatically conserved, i.e. $\partial_\mu k^\mu=0$, and the topological "magnetic" charge is defined as $g_{M}=\int d^{3} x k_{0}$. Following Ref.~\cite{shnir}, the integration can be carried out, going from a volume integral to a surface integral over the sphere $S^2$ at infinity. Parametrizing local coordinates on $S^2$ by $\sigma_\alpha$, $\alpha=1,2$, one obtains \cite{shnir}
\be
\label{gMint}
g_{M}=\int d^{3} x k_{0}=\frac{1}{g}\int d^2 \sigma \sqrt{{\bf g}} \,,
\ee 
with ${\bf g}=det(\partial_\alpha \hat{\xi}^a \partial_\beta \hat{\xi}^a)$ and $\hat{\xi}^a \equiv \xi^a/v_M$. With $\int d^2 \sigma \sqrt{{\bf g}} = 4 \pi n$, where $n$ is the so-called Brower index, one obtains the following topological quantization condition

\be
\label{quantize}
\frac{g g_{M}}{4\pi}=n \,,
\ee                       
Notice that the quantization condition (\ref{quantize}) is in terms of the monopole topological charge $g_{M}$ and the {\em weak charge} $g$ instead of the usual $e$ of the Dirac quantization condition. In fact, the unit magnetic charge, for $n=1$, is given by (using $g=e/\sin \theta_W$)
\be
\label{unit}
g_{M}^{(0)} = \frac{4\pi}{g}=(\frac{4 \pi}{e})\sin \theta_W \,,
\ee
\be
\label{higher}
g_{M}^{(n)}=n\, g_{M}^{(0)}; n \geq 2 \,.
\ee
To paraphrase Goddard and Olive \cite{G-O}, the magnetic charges are topologically conserved and quantized in units of $(\frac{4 \pi}{e})\sin \theta_W$.


Let us notice that the quantization used here and elsewhere in the literature is done in the quantum field theory framework where $c=1$ and, hence, there is the appearance of the factor $4 \pi$ in the denominator of the left-hand-side of Eq.~(\ref{quantize}). To get to the familiar form of the Dirac quantization, one redefines $(g,g_{M})/\sqrt{4\pi} \rightarrow (g,g_{M})$ giving 
\be
\label{quantize2}
g\, g_{M}=n \,. 
\ee
With that convention, our unit magnetic charge becomes 
\be
\label{unit2}
g_{M}^{(0)} = \frac{1}{g}=(\frac{1}{e})\sin \theta_W \,.
\ee 

It has been shown by \cite{EHM} that, far from the core of the monopole, the requirement that the wave function of a charged particle such as the electron is single-valued as it goes around the monopole imposes an extra {\em Dirac quantization condition} 
\be
\label{Dirac}
e g_{M}=m/2 \,.
\ee
For a topological unit magnetic charge, combining Eq.~(\ref{unit2}) with Eq.~(\ref{Dirac}), one obtains 
\be
\label{sinw}
\sin^{2} \theta_W = \frac{m^2}{4} \,.
\ee
Ref.~\cite{EHM} has argued that only $m=1$ is possible since $\sin^{2} \theta_W <1$ leading to a very interesting prediction
\be
\label{sinw2}
\sin^{2} \theta_W = \frac{1}{4} \,.
\ee
It is interesting to note that, if the integer $m$ coming from the Dirac quantization condition is identified with the topological integer $n$, one would arrive at the same prediction (\ref{sinw2}) for {\em any}  $m$. As shown in Ref.~\cite{EHM}, the value of $\sin^{2} \theta_W = \frac{1}{4}$ at the monopole mass becomes, from a leading-order renormalization-group analysis, ${\rm sin}^2 \theta_W \simeq 0.231$ at the $Z$-boson mass, in agreement with experiment. With the assumption $m=n$, the prediction $\sin^{2} \theta_W = \frac{1}{4}$ is valid for all $n$ and $m$ and it would imply the topological nature of the electroweak monopole, in contrast with the point-like Dirac monopole,.

3) One of the most important results of the present manuscript is the estimate of the mass of the monopole and the size of its core. 
\begin{itemize}
\item The calculation of that mass is well known. What is different here from the usual estimates is the value of the VEV of the real triplet which is less than the electroweak scale. One has

\be
\label{mass}
M=\frac{4\pi v_M}{g} f(\lambda/g^2) \,,
\ee
where once again, in order to avoid confusion, $g$ is the $SU(2)$ gauge coupling and $\lambda$ is the $\xi$ self-coupling. It is well known that the function $f(\lambda/g^2)$ varies between 1 for $\lambda=0$ (Prasad-Sommerfield limit) and 1.78 for $\lambda=\infty$. From Ref.~(\cite{pqnur}), the value of $v_M$ is constrained from below by the Z-width (only three light neutrinos) i.e. $v_M > M_Z/2 \sim 45.5 \gev$ and from above by $(\sum_{i=1,2} v_i^2 + v^{M,2}_{i}) + 8 v_M^2 = (246 \gev)^2$. Taking the lowest value corresponding to $v_M \sim 45.5 \gev$ and $f=1$ and the largest value corresponding to $v_M \sim 87 \gev$ and $f=1.78$, we obtain the following bound for the monopole mass
\be
\label{massbound}
889 \gev \alt M  \alt 2993 \gev \,.
\ee

Notice that the monopole mass is proportional to the triplet VEV $v_M$ and so is the right-handed Majorana mass $M_R$ of $\nu_R$. The search for $\nu_R$ is intrinsically linked to the search for the monopole.
It goes without saying that the above estimate is for the sole purpose of showing that the monopole mass $M \sim O(\tev)$ and is, therefore, {\em accessible experimentally}. This mass is {\em much smaller}, by at least {\em thirteen orders of magnitude}, than a typical mass of a GUT monopole.

\item The monopole has a core of radius $R_c \sim (gv_M)^{-1} \sim 10^{-16} cm$, roughly a thousand times {\em smaller} than a proton "radius". 
Here one has $g v_M$ appearing in the denominator since that is the contribution to the W-boson mass from $\xi$. Inside the core are virtual $W^{\pm}$ and $Z$. Far from the core, this monopole behaves like a Dirac monopole with {\em some caveat} as we shall see below.

In summary, due to the presence of the real triplet $\xi$ of $SU(2)$, a topologically stable monopole exists as a finite-energy soliton with finite size core (no singularity as opposed to a Dirac monopole) and with a mass of $O(\tev)$.

\end{itemize}

5) The next step is to look at the interaction of the perturbative spectrum of the full $SU(2)_W \times U(1)_Y$ with the monopole i.e. small fluctuations of these fields in the presence of the monopole background. This procedure is reviewed in detail in an excellent review of Shnir \cite{shnir} and we will adopt it here. (It is not needed for the purpose of this manuscript but it is included here for future investigations.)
We will denote the {\em small fluctuations} as $w_{\mu}^a$ and $b_{\mu}$ for the gauge fields, and $\xi$, $\chi$, $\phi^{SM}_{i}$ and $\phi^{M}_{i}$ ($i=1,2$) for the scalars. 

Let us start with the pure $SU(2)$ case with the real triplet $\xi$. We have the following modifications for Eq.~(\ref{ansatz}):
\be
\label{fluctuation}
\tilde{\xi}^a=\Xi^a + \xi^a; \tilde{W}_{n}^{a}=W_{n}^{a} + w_{n}^{a};W_{0}^{a}=w_{0}^{a}.
\ee
As it has been discussed in details in \cite{shnir}, the discussion of the dynamical equations governing the aforementioned small fluctuations is carried out in the simplest way in the {\em Unitary gauge}. Notice that, if we had only $\xi$ which carries zero $U(1)_Y$ quantum number, its VEV would induce $SU(2)_W \times U(1)_Y \rightarrow U(1)_W \times U(1)_Y$. Let us first study this step before the $U(1)_W \times U(1)_Y$ symmetry is broken down to $U(1)_{em}$ by other Higgs fields.  It is beyond the scope of this note to repeat well-known results for this case. Here, we just quote the essential points which can be found in \cite{shnir}.

We will not just look at the perturbative spectrum of $SU(2)$ with a real scalar triplet but with the entire scalar spectrum, including a complex triplet $\chi$ and four doublets $\Phi^{SM}_{i}$ and $\Phi^{M}_{i}$. Furthermore, the full gauge group is $SU(2) \times U(1)_Y$.

From 
\be
\label{Fmunu}
W^{3}_{\mu \nu} = \vec{\xi}.\vec{W}_{\mu \nu}/v_M \,,
\ee 
the familiar results for {\em static} field strengths are obtained
\be
\label{Fij} 
W^{3}_{ij} = \frac{\epsilon_{ijk} \hat{\vec{x}}^k}{g r^2} \,,
\ee  
and
\be
\label{B}
B_{i} = -\frac{1}{2} \epsilon_{ijk} W^{3,jk}=\frac{1}{g r^2} \hat{r}_{i} \,,
\ee  
a well-known static radial magnetic field. Outside the core of radius $R_c \sim (gv_M)^{-1}$, one has a long-range magnetic field of strength {\em $1/g$}.


We now include the other Higgs fields: the complex triplet $\tilde{\chi}$ and the doublets, which spontaneously break $SU(2)_W \times U(1)_Y$ down to $U(1)_{em}$. In the topological language, this means that we now include the spaces $S^{5} \times \sum_{i=1,2} S^{3}_{SMi} \times \sum_{i=1,2} S^{3}_{Mi}$ and the vacuum manifold is described by $S_{vac}$ as shown in Eq.~\ref{vacuum}. As we have shown above, this vacuum manifold can still support the existence of a topologically stable monopole in light of Eq.~(\ref{result}). However, $W_{\mu}^{3}$ is no longer a mass eigenstate but is now written in terms of the Z-boson and photon fields as $W_{\mu}^{3}= \cos \theta_W Z_{\mu} + \sin \theta_W A_{\mu}$. As a result, one has
\be
\label{W3new}
W^{3}_{ij} =  \cos \theta_W Z_{ij} + \sin \theta_W F_{ij} \,,
\ee  
where $F_{ij}$ is the usual electromagnetic field strength tensor and $Z_{ij}$ is the Z field strength tensor. This is the reason why the name "$\gamma$-Z magnetic monopole " was chosen. Since $Z_{ij}$ contains the static "electric" and "magnetic" Z fields, it has an exponential damping factor $\exp(-M_{Z} r)$ where $M_Z$ denotes the Z-boson mass. One can now generalizes Eq.~(\ref{B}) as
\bea
\label{gammaZ}
B^{\gamma Z}_{i}&=&  \frac{1}{g r^2} \hat{r}_{i} ( \cos \theta_W e^{-M_{Z} r} + \sin \theta_W ) \\ \nonumber
                            &=& \frac{\sin \theta_W}{e r^2} \hat{r}_{i} ( \cos \theta_W e^{-M_{Z} r} + \sin \theta_W ) \,,
\eea  
where $e$ appearing in (\ref{gammaZ}) denotes the usual electromagnetic coupling and where we have used the usual SM relationship $e= g \sin \theta_W$. Away from the core, the true magnetic field is simply
 \be
\label{Bnew}
B_{i}  \approx  \frac{\sin \theta_W}{e r^2} \hat{r}_{i} \,.
\ee    
A few remarks are in order here. First, in the limit that the VEVs of all Higgs fields {\em except} for $\xi$ vanish, $M_Z =0$, $\theta_W=0$, $B^{\gamma Z}_{i} \rightarrow B_{i}$ and one recovers the 't Hooft-Polyakov result. Second, from Eq.~(\ref{gammaZ}), one notices that, at large distances $r \gg R_c$, the magnetic field differs in strength from that of a point-like Dirac monopole by a factor $\sin \theta_W$. Third, the short-range and long-range parts of $B^{\gamma Z}_{i}$ become comparable in strength at a distance 
$r= \frac{1}{M_Z} \ln (\cot \theta_W) \sim 0.6/M_Z$. This is well inside the core of the monopole.  A summary of the properties of the $\gamma$-Z magnetic monopole is in order here.
 
 1) The existence in the EW-$\nu_R$ model of a real Higgs triplet $\xi$ gives rise to topologically-stable, finite-energy electroweak monopole;
 
 2) The monopole mass, $M=\frac{4\pi v_M}{g} f(\lambda/g^2) \sim 889 \gev - 2.993 \tev$ is intrinsically linked to the Majorana masses of the right-handed neutrinos.
 
 3) The monopole is a finite-energy soliton with a core of radius $R_c \sim (gv_M)^{-1} \sim 10^{-16} cm$, with virtual $W^{\pm}$ and $Z$ inside the core.
 
 4) This electroweak monopole has a long-range magnetic field $B_{i}  \approx  \frac{\sin \theta_W}{e r^2} \hat{r}_{i}$ at distances larger than the core radius. 
 As shown in Ref.~\cite{EHM}, this particular aspect of the electroweak monopole leads to the interesting prediction of $\sin^{2} \theta_W$ outside the framework of Grand Unified Theories and in agreement with experiment.
 
 5) For generalization of the theory, besides the monopole solution with $W_0^a=0$, it may be worth to add the dyon solution with $W_0^a=\frac{r^a}{g r^2} J(r) \neq 0$ which carries both magnetic and electric charges. It is beyond the scope of this paper to discuss such a possibility. It will be treated elsewhere.
 
 
\section{Production and detection of the $\gamma$-Z magnetic monopole} 
  
  We will make a few remarks in this section and postpone a more detailed treatment of the search for the $\gamma$-Z magnetic monopole to a longer version. This could be considered to be a very brief overview of the search for monopoles.
  
  The first thing to notice is the coupling strength of the monopole $g_M$. 
 This is the same as the usual estimate for a point-like Dirac monopole $g_D =e/2\alpha_{em} \approx 68.5 \, e$. Because of the fact that the coupling is large, there is a large uncertainty in the calculation of the production cross section of $M\bar{M}$, where $M$ stands for a monopole. It goes without saying that a non-perturbative treatment is needed. It is useful nevertheless to get some rough ideas about what one might expect from colliders such as the LHC.
  
 There are great uncertainties in making an estimate for the production cross section. In general, this estimate is highly non-perturbative. In addition, it has been argued that the production process depends on the initial states which could be, for instance, p-p collisions at the LHC or heavy-ion collisions, also at the LHC. In the first case of p-p collisions, arguments have been given for why the cross section for the production of a pair of composite 't Hooft-Polyakov-like monopoles is {\em exponentially-suppressed} as $\sigma \sim \exp(-4/\alpha =-548)$ and thus ruling out the production of such monopoles even if they are "light" enough to be pair-produced \cite{nussinov}. In a nutshell, the argument implies that the energy carried by a few degrees of freedom of the initial p-p states has to be distributed among a large number O($1/\alpha$) of coherent states that the composite monopoles carry (unlike the point-like Dirac monopoles) and hence an exponential suppression factor in the cross section. One may, however, suspect that such an argument might be insufficient considering the non-perturbative nature of the production process
 and is not a definite proof. It goes without saying that a more comprehensive analysis of this topic is needed. An alternative proposal \cite{gould} was to use heavy-ion collision because the production process is very different from that of a p-p collision, coming mainly from a thermal Schwinger thermal pair production process. It has been argued that this process is valid for both cases of point-like Dirac monopoles or soliton 't Hooft-Polyakov-like monopoles. Taking into account the aforementioned caveat, let us go ahead and NAIVELY estimate what most likely is an upper bound on the expected number of monopoles of a given mass at the LHC.
   
 Leading-order calculations for the production cross sections through Drell-Yan and $\gamma \gamma$ fusion at the 13 GeV LHC have been carried out \cite{moedal}. 
 A rough estimate for a 2-TeV monopole production cross section is $\sim 40 fb$, assuming a spin-0 monopole and assuming it is point-like. It is obvious that this estimate is for the sole purpose of illustration. With the projected luminosity of the high luminosity LHC (HL-LHC) to be around $250 fb^{-1}$/year, one might expect $10,00$ monopole events. These numbers are, of course, to be taken with a BIG GRAIN OF SALT. As mentioned above, heavy-ion collisions are another venue that one can explore.
 
 There have been extensive discussions on the various methods of detection of electroweak-scale monopoles. These include highly ionizing tracks in detectors such as ATLAS or LHCb. Monopoles could be trapped by Al nuclei in the MoEDAL detector \cite{moedal} or the Beryllium CMS beam pipe. It goes without saying that this kind of search is sufficiently important to motivate additional novel techniques.
 
 \section{Conclusion}
 
 The construction of a topologically stable, finite-energy electroweak monopole is first proposed within the framework of a model (the EW-$\nu_R$ model \cite{pqnur}) in which right-handed neutrinos are non-sterile and have masses proportional to the electroweak scale $\Lambda_{EW} \sim 246 \gev$. To ensure that Custodial Symmetry is preserved ($M_W = M_Z  \cos \theta_W$ at tree-level), the model requires the existence of a {\em real} Higgs triplet (in addition to a complex Higgs triplet which gives Majorana masses to the right-handed neutrinos) whose VEV is proportional to $\Lambda_{EW}$. It is this real Higgs triplet which gives rise to the aforementioned electroweak monopole. The magnetic charge of this monopole is topological and whose smallest value is found to be $g_M=1/g$ where $g$ is the $SU(2)_W$ coupling.
 
 This monopole is a finite-energy soliton with a core of radius $R_c \sim (gv_M)^{-1} \sim 10^{-16} cm$. Its mass is estimated to lie between $ 900 \gev$ and $3 \tev$ and it is intrinsically linked to the Majorana masses of the right-handed neutrinos. It could be searched for at various LHC detectors such as ATLAS, LHCb, CMS, and, in particular, at the dedicated experiment MoEDAL where it could be trapped by Aluminum nuclei in its detector.
 
 On the theoretical front, it was realized \cite{EHM} that the imposition of the Dirac Quantization Condition led to a prediction for $\sin^{2} \theta_W=1/4$ at the electroweak mass scale which becomes, from a leading order renormalization group analysis, ${\rm sin}^2 \theta_W \simeq 0.231$ at the $Z$-boson mass. This prediction outside the framework of Grand Unified Theories is remarkable in its own right.
 
 An interesting generalization of the model, besides the monopole solution with $W_0^a=0$, is to add the dyon solution with  $W_0^a=\frac{r^a}{g r^2} J(r) \neq 0$ \cite{EHM2}. Such an interesting object could also be searched for at MoEDAL.
 
 The presence of a "low scale" electroweak monopole suggests that a Rubacov-Callan-type. fermion catalysis \cite{RC} might be accessible experimentally. Furthermore, could baryogenesis be generated in the presence of electroweak monopoles? Could electroweak monopole bound states be candidates for dark matter as suggested by \cite{csaki} for a particular type of TeV monopole? These questions are under investigation.

%
%
%
%
%
%
%

\end{document}